# IMAGE ENCRYPTION USING FIBONACCI-LUCAS TRANSFORMATION

Minati Mishra[1], Priyadarsini Mishra[2], M.C. Adhikary[3] and Sunit Kumar[4]

[1]P.G. Department of Information & Communication Technology, F.M. University, Balasore, Odisha, India
[2]District Rural Development Agency, Balasore, Odisha, India
[3]Department of Applied Physics and Ballistics, Fakir Mohan University, Balasore, Odisha, India
[4]Jamshedpur Co-operative College, Kolhan University, Jharkhand, India

## ABSTRACT

*Secret communication techniques are of great demand since last 3000 years due to the need of information security and confidentiality at various levels of communication such as while communicating confidential personal data , patients' medical data, countries'  defence and intelligence information, data related to examinations etc. With advancements in image processing research, Image encryption and Steganographic techniques have gained popularity over other forms of hidden communication techniques during the last few decades and a number of image encryption models are suggested by various researchers from time to time. In this paper, we are suggesting a new image encryption model based on Fibonacci and Lucas series.*

## 1. INTRODUCTION

With rapid advancement in Internet and networking technologies during the recent years, communication and information exchange have become much easier and faster but at the same time the issues related to data security and confidentiality have become a major concern of the time. To cater to this need of information security, a number of hidden and secret communication techniques such as cryptography, anonymity, covert channels, Steganography, Watermarking etc have been developed. Out of all these methods, the digital image Steganographic methods have been heavily used by the researchers during the last few decades for the purpose of secret communication and information authentication due to the size and popularity of digital images. The digital image Steganographic methods generally depends upon various image scrambling techniques in order to further improve the level of security of the hidden information. Image scrambling techniques scramble the pixels of an image in such a manner that the image becomes chaotic and indistinguishable. These scrambling techniques generally use several keys for encryption and decryption and without the correct keys and an appropriate method; the third party users cannot access the secret information even if they are able to sniff the medium. Hence, the message remains highly secured against unauthorised access. Even though, a number of image scrambling techniques have been developed by different researchers during the last two decades, a lot of research is still going on in this area. Here in this paper we have developed a simple but powerful 2x2 chaotic map combining the most famous Fibonacci and Lucas series.  Just like the other popular chaotic maps such as Arnold's cat map and Fibonacci-Q, this is also a periodic map but unlike these two methods which use a single 2x2 map, this proposed method has a number of 2x2 maps and hence, this method provides higher security over the previous methods.

## 2. IMAGE SCRAMBLERS

Due to the all time need for data security and confidentiality, image scrambling methods have emerged as popular encryption standards during last four decades and many researchers have worked in this filed to suggest a number of such methods by now. The most primitive and powerful method in this category was proposed by Vladimir Arnold in 1960 [1], [2]. The journey led by Arnold during 1960 has been followed and redefined by several researchers during last two decades giving rise to several interesting image transformation techniques. Here is a brief review of some such methods.

Ma and Qiu developed a cryptosystem using the general cat map in 2003[3]. Kong and Dan in 2004 developed a new Anti-Arnold algorithm [4]. Hong and Zou extended the 2D Arnold transform to 3D and also studied the period of the Arnold transform (AT) [5]. Wang in his paper made a study on the period of 2D random matrix scrambling transform and used it for image hiding [6]. Yang et al. in 2006 gave a digital image scrambling technology based on the symmetry of AT [7]. Minati Mishra et al. in their papers further extended the method and generalised the AT to improve the security of the scrambled information [8], [9]. When a group of people were concentrating on AT, at the same time, Qi et al in 2000 gave a new class of transform and its application in the image transform covering [10] which is further extended by Zou et al. ,who have used Fibonacci numbers and generalised Fibonacci transform to scramble image in spatial domain [11], [12]. Li-Ping Shao et al in their paper studied 2D Triangular Mappings and Their Applications in Scrambling Rectangular Images [13]. Many other researchers from time to time used one or more of these above scrambling methods, combined these methods with many other frequency domain transform methods in their works on Steganography and watermarking. In this paper we are limiting our study to the two dimensional spatial domain scramblers only.

### 2.1 Arnold Transform

**Definition:** It is a transformation $\Gamma : T^2 \to T^2$ such that:

$$\begin{pmatrix} x' \\ y' \end{pmatrix} = \begin{pmatrix} 2 & 1 \\ 1 & 1 \end{pmatrix} \begin{pmatrix} x \\ y \end{pmatrix} (\mod \quad N) \tag{1}$$

Where, x, y ∈ {0, 1, 2 … N −1} and N is the size of a digital image.

A new image is produced when all the points in an image are manipulated once by equation (1). This is a simple but powerful transform [1], [2], [3] which is periodic in nature and is very much popular in spatial domain applications.

### 2.2 Modified Arnold Transform

The security level of the encrypted text becomes low when it is encrypted using the basic Arnold transform as it constitutes a single 2x2 map and therefore, can easily be decrypted by any 3[rd] party user, using the same map. To enhance the security level of the encrypted message the basic AT is generalised [8], [9] as follows:

$$\begin{pmatrix} x' \\ y' \end{pmatrix} = \begin{pmatrix} k+1 & k \\ 1 & 1 \end{pmatrix} \begin{pmatrix} x \\ y \end{pmatrix} (\mod N) \qquad (2)$$

OR

$$\begin{pmatrix} x' \\ y' \end{pmatrix} = \begin{pmatrix} k & k+1 \\ 1 & 1 \end{pmatrix} \begin{pmatrix} x \\ y \end{pmatrix} (\mod N) \qquad (3)$$

Where, x, y ∈ {0, 1, 2 … N −1} and N is the size of a digital image and k ∈ {0, 1, 2, 3…}.

As in both the cases the transform matrices are 2x2 unimodular matrices therefore, are periodic in nature and scramble a square image into an indistinguishable format. Unlike equation (1) where there is a single map, equation (2) and (3) provides a number of maps for different values of k and hence increases the security level of the scrambled message against hit and trial decryption by an unauthorised user. In fact it can be easily be seen that the transposes of the 2x2 maps given in equations (2) and (3) and the matrices obtained by swapping the rows as well as their transposes; all the 8 variants of $\begin{pmatrix} k+1 & k \\ 1 & 1 \end{pmatrix}$ are generalisations of the AT and can be used for the purpose of image encryption. For k=1, equation (2) is same as the basic AT of equation (1).

### 2.3 Fibonacci-Q Transform

Qi et al during 2000 used Fibonacci-Q transformation replacing 2D Arnold transformation matrix by 2D Fibonacci- Q transformation [10]. This is a special case of the basic AT with periodicity. The equation of the transformation is as given below:

$$\begin{pmatrix} x' \\ y' \end{pmatrix} = \begin{pmatrix} 1 & 1 \\ 1 & 0 \end{pmatrix} \begin{pmatrix} x \\ y \end{pmatrix} \mod N \qquad (4)$$

Li-Ping Shao et al further worked on the set of triangular periodic transforms and used those for image scrambling in their paper [13]. The generalised triangular map is given by $\begin{pmatrix} 0 & 1 \\ 1 & k \end{pmatrix}$, k∈ {0,1,2…}and the other three possible variants of this which can be obtained by rearranging the values of the matrix. Though these are the simplest forms of 2D transforms, one important limitation of all these triangular matrices is, one out of the two coordinates of the image points always remains constant and only one coordinate changes during the iterations making the scrambling pattern less random.

### 2.4 Generalised Fibonacci Transform

Named after Leonardo of Pisa, popularly known as Fibonacci, the Fibonacci sequence $F_n$, is a sequence of integers given by the recurrence relation

$$F_n = \begin{cases} 0, & if \quad n=1 \\ 1, & if \quad n=2 \\ F_{n-1} + F_{n-2} & otherwise \end{cases} \qquad (5)$$

The series constitutes the numbers:

$$0,1,1,2,3,5,8,13,21, 34..$$

It can be easily seen that a 2x2 matrix formed by any four consecutive terms of the Fibonacci series is a unimodular matrix and can be considered as an image scrambler. A generalised Fibonacci Transform is defined as:

**Definition:** The generalized form of the Fibonacci Transform is a mapping $F : T^2 \to T^2$ such that:

$$\begin{pmatrix} x' \\ y' \end{pmatrix} = \begin{pmatrix} F_i & F_{i+1} \\ F_{i+2} & F_{i+3} \end{pmatrix} \begin{pmatrix} x \\ y \end{pmatrix} (\mod N) \qquad (6)$$

Where, x, y ∈ {0, 1, 2 … N −1}, Fi is the i$^{th}$ term of the Fibonacci series and N is the size of a digital image.

Denoting $\begin{pmatrix} F_i & F_{i+1} \\ F_{i+2} & F_{i+3} \end{pmatrix}$ as FT$_i$, the first matrix of this series will be given by:

$$FT_1 = \begin{pmatrix} F_1 & F_2 \\ F_3 & F_4 \end{pmatrix} = \begin{pmatrix} 0 & 1 \\ 1 & 2 \end{pmatrix} \qquad (7)$$

And this way we can form many Fibonacci transforms [11], [12] for different values of i. Just like the modified Arnold Transforms, all of these maps will be periodic in nature with a maximum possible periodicity $N^2$-1. Like the modified AT, in this case too there are a number of different maps with different periodicities and different scrambling patterns and hence these transforms also, just like the Modified ATs, can be considered as more secured encryption methods over the basic AT and Fibonacci-Q transform.

## 3. PROPOSED FIBONACCI- LUCAS TRANSFORM

### 3.1 Lucas Series

Named after the French mathematician François Édouard Anatole Lucas, who has studied both the Fibonacci series and Lucas series, Lucas series is a special case of Fibonacci series and is defined by the recurrence relation:

$$L_n = \begin{cases} 2, & if \quad n = 1 \\ 1, & if \quad n = 2 \\ L_{n-1} + L_{n-2} & otherwise \end{cases} \qquad (8)$$

The series constitutes the numbers:

$$2, 1, 3, 4, 7, 11, 18, 29… \qquad (9)$$

Unlike the Fibonacci series, the terms of the Lucas series does not form a unimodular periodic map and therefore, cannot be used for image encryption but by combining the terms of the Fibonacci and Lucas series we can form a series of new periodic maps those can be used for image scrambling.

### 3.2 Fibonacci- Lucas Transform

Several special cases of the Fibonacci series can be constructed by changing the seed values. For example, considering the seed values to be 1 and 1 instead of 0 and 1, one variant of the Fibonacci series becomes:

$$1,1,2,3,5,8,13,21, 34... \qquad (10)$$

Let us denote this above series as Fibo11 (the postfix 11 to indicate that the seeds are 1 and 1) series. In a similar way we can construct other variants of the series taking the initial values 3, 2 and 3, 1. Denoting these variants of the Fibonacci series as Fibo32, and Fibo31 respectively, the elements of these series will be given by:

$$\begin{aligned} \text{Fibo32} &= 3, 2, 5, 7, 12, 31, 50, 81, 131\ldots \\ \text{Fibo31} &= 3, 1, 4, 5, 9, 14, 23, 37\ldots \end{aligned} \qquad (11)$$

Combining the terms of Fibo11 or Fibo32 or Fibo31 series with the corresponding terms of the Lucas series, we can define a set of new image transformations as follows:

**Definition:** The Fibonacci-Lucas Transform can be defined as the mapping $FL: T^2 \to T^2$ such that:

$$\begin{pmatrix} x' \\ y' \end{pmatrix} = \begin{pmatrix} F_i & F_{i+1} \\ L_i & L_{i+1} \end{pmatrix} \begin{pmatrix} x \\ y \end{pmatrix} (\mod \ N) \qquad (12)$$

Where, x, y ∈ {0, 1, 2 … N −1}, $F_i$ is the $i^{th}$ term of the Fibo11 series and $L_i$ is the $i^{th}$ term of the Lucas series, (i = 1, 2… except for i=3),   N is the size of a digital image.

Denoting $\begin{pmatrix} F_i & F_{i+1} \\ L_i & L_{i+1} \end{pmatrix}$ as $FLT_i$, the first matrix of this series will be given by:

$$F(11)LT_1 = \begin{pmatrix} F_1 & F_2 \\ L_1 & L_2 \end{pmatrix} = \begin{pmatrix} 1 & 1 \\ 2 & 1 \end{pmatrix} \qquad (13)$$

Continuing in this way we can form an infinitely many transforms and just like the Modified Arnold and Fibonacci Transforms, all of these matrices will be periodic in nature with a maximum possible periodicity of $N^2-1$ and will produce scrambling patterns different from each other.

The first 18 terms of the specialised Fibonacci series (equations (10) and (11)) and the Lucas series (equation (9)) are given below in table1:

Table 1. Terms of Fibonacci and Lucas series

| n | 1 | 2 | 3 | 4 | 5 | 6 | 7 | 8 | 9 | 10 | 11 | 12 | 13 | 14 | 14 | 16 | 17 | 18 |
|---|---|---|---|---|---|---|---|---|---|----|----|----|----|----|----|----|----|----|
| F11$_n$ | 1 | 1 | 2 | 3 | 5 | 8 | 13 | 21 | 34 | 55 | 89 | 144 | 233 | 377 | 610 | 987 | 1597 | 2584 |
| L$_n$ | 2 | 1 | 3 | 4 | 7 | 11 | 18 | 29 | 47 | 76 | 123 | 199 | 322 | 521 | 843 | 1364 | 2207 | 3571 |
| F32$_n$ | 3 | 2 | 5 | 7 | 12 | 19 | 31 | 50 | 81 | 131 | 212 | 343 | 555 | 898 | 1453 | 2351 | 3804 | 6155 |
| F31$_n$ | 3 | 1 | 4 | 5 | 9 | 14 | 23 | 37 | 60 | 97 | 157 | 254 | 411 | 665 | 1076 | 1741 | 2817 | 4558 |

### 3.3 Periodicity of Fibonacci-Lucas Transformation

From the above table 1, it is obvious that $L_i=F11_i+F11_{i-2}$, for all $i>=3$. Substituting this value of $L_i$ in equation (4), $FLT_i$ can be rewritten as:

$$\begin{pmatrix} F11_i & F11_{i+1} \\ F11_i + F11_{i-2} & F11_{i+1} + F11_{i-1} \end{pmatrix}. \tag{14}$$

Using method of induction, it can be easily shown that:

$$\det(FLT_k) = (-1)^k \text{ for all } k>=4 \tag{15}$$

This shows that the 2x2 maps formed taking any two consecutive terms of the Fibonacci series and the corresponding two terms of the Lucas series from the above table from the 4th term onwards form periodic maps. Moreover, it can be seen that except for the red lettered values, for all other terms of the above table-1 form periodic maps and can safely be used for the purpose of image encryption. In fact Lucas series is such an interesting series that this series in combination with Fibo32 and Fibo31 also produces an infinite number of unimodular periodic transforms those can be used for image encryption.

## 3. EXPERIMENTAL RESULTS AND DISCUSSIONS

### 3.1 Periodicity

The transformation periods of the Generalised Fibonacci Transform (GFT), Modified/ Generalised Arnold Transform (GAT) [8] and the Fibo-Lucas Transform (FLT) for the 128x128 gray scale Lena image are given in the following table 2:

Table 2. Periods of Fibonacci-Lucas and other transforms for 128x128 gray scale Lena Image for different values of i

| i | 1 | 2 | 3 | 4 | 5 | 6 | 7 | 8 | 9 | 10 | 11 | 12 | 13 | 14 | 15 | 16 |
|---|---|---|---|---|---|---|---|---|---|----|----|----|----|----|----|----|
| **GFT** | 128 | 64 | 128 | 128 | 16 | 128 | 128 | 64 | 128 | 128 | 8 | 128 | 128 | 64 | 128 | 128 |
| **GAT** | 128 | 192 | 64 | 192 | 128 | 192 | 32 | 192 | 128 | 192 | 64 | 192 | 128 | 192 | 16 | 192 |
| **F(11)LT** | 128 | 64 | 128 | 128 | 16 | 128 | 128 | 64 | 128 | 128 | 8 | 128 | 128 | 64 | 128 | 128 |
| **F(32)LT** | 64 | 96 | 192 | 32 | 192 | 96 | 64 | 12 | 192 | 32 | 192 | 48 | 64 | 96 | 192 | 32 |
| **F(31)LT** | 64 | 64 | 32 | 64 | 64 | 8 | 64 | 64 | 32 | 64 | 64 | 4 | 64 | 64 | 32 | 64 |

It is clear from the above table that the generalised Fibonacci transform, as given by equation (6) and the Fibo-Lucas transform given in equation (12) and the generalised/ modified AT (equation 3) vary from each other for different values of i. The periods of the basic AT (equation 1), Modified AT (for k=1, equation 3), Fibonacci-Q transform (equation 4), $FT_1$ (equation 4) and $F(11)LT_1$(equation 12), $F(32)LT_1$, $F(32)LT_1$ for the same 128x128 gray scale Lena image are as given in table 3.

Table 3. Periods of various Transforms

| Transform | Basic Arnold | Generalised Arnold for k=1, eq.3 | Fibonacci-Q | FT1 | F(11)LT1 | $F(32)LT_1$ | $F(31)LT_1$ |
|---|---|---|---|---|---|---|---|
| **Period** | 96 | 128 | 192 | 128 | 128 | 64 | 64 |

## 3.2 Scrambling patterns

Let,  A = [1 2 3; 4 5 6; 7 8 9]  (15)

be a 3x3 matrix. Table 4 lists the matrices after transforming A for three times each and Figure. 1 shows the scrambled Lena image after scrambling it 25 times by the methods given in table 3.

Table 4. Scrambled matrices of A after 3 times scrambled by various Transforms

| Transform | Basic Arnold | Generalised Arnold for k=1, eq.3 | Fibonacci-Q | FT1 | FLT1 |
|---|---|---|---|---|---|
| Period | 4 | 8 | 8 | 8 | 8 |
| Three times transformed | 1 5 9<br>8 3 4<br>6 7 2 | 1 6 8<br>9 2 4<br>5 7 3 | 1 7 4<br>9 6 3<br>5 2 8 | 1 8 6<br>3 7 5<br>2 9 4 | 1 9 5<br>8 4 3<br>6 2 7 |

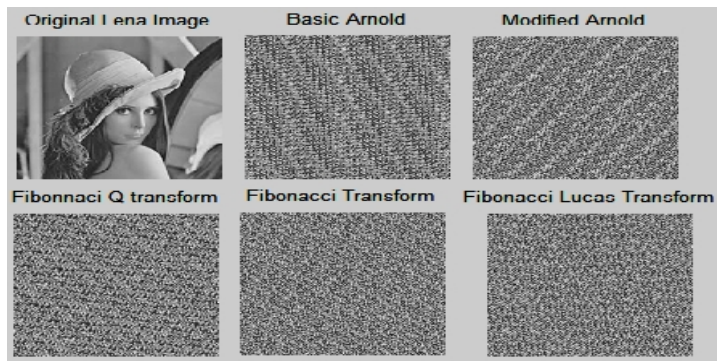

Figure 1. 25times Transformed Lena Image

## 3.3 Robustness against statistical attacks

To test the robustness of the method against various statistical attacks, we have tried to decrypt the scrambled images those are compressed using the standard baseline JPEG compression with quality factors up to as low as 4 and could successfully recover the image. It is also found that the method is robust against random cropping, format change and noise introduction. The results of our experiments are given below in figure 2, 3 and 4. Figure 2 shows the recovered images after the image is scrambled 30 times, saved in TIFF format and then changing the format to PNG, BMP and JPEG. Figure 3 shows recovery against compression with different quality factors and cropping. Figure 4, shows the recovered images after the encrypted image is introduced with various types of standard noises such as Salt & Pepper, Gaussian, speckle with different densities and variances. From all these above results, it is clear that the encryption process is very much robust against compression, cropping, noise and format change and the image can be recovered without much loss of image information. Of course, the method failed to recover the image against rotation and scaling.

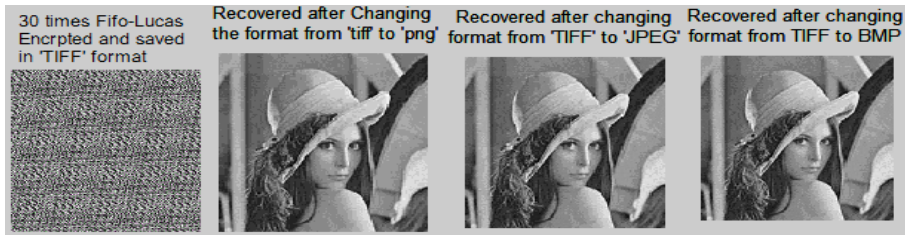

Figure 2. Recovery after format change

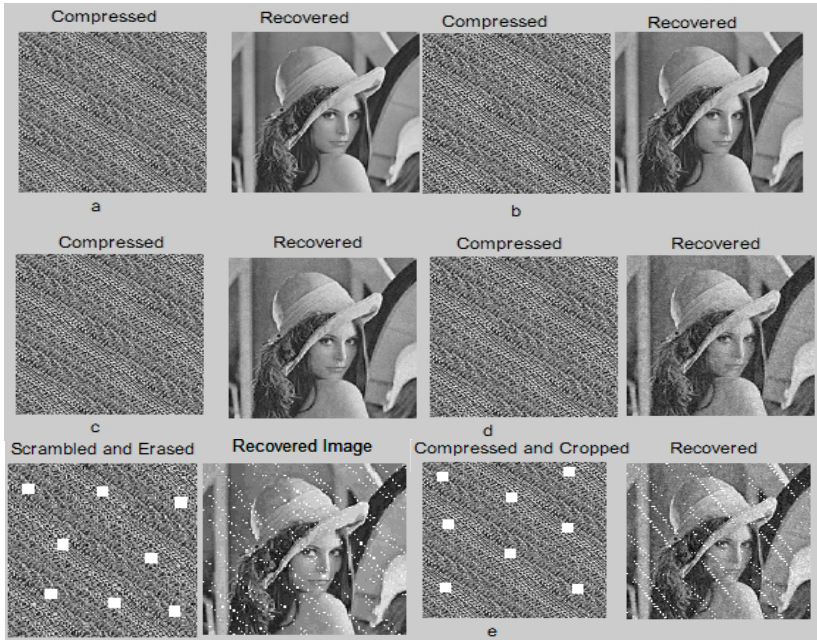

a, b, c, d : Baseline Standard JPEG compressed images with quality factors 10, 8, 6, 4 respectively.
e: cropped d

Figure 3. Recovery against crop and compression attacks

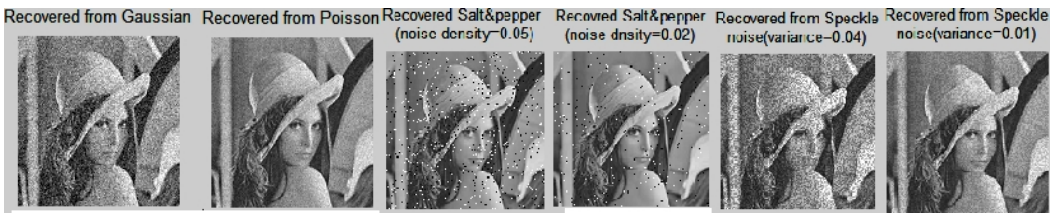

Figure 4. Image recovered from various noise attacks

### 3.4 Recommendations based on observations and findings

Observation 1: It is noticed that for the transforms $GAT_i$, $GFT_i$ and $FLT_i$, values of 'i' for which the transforms attain maximum periodicity (i. e., $N^2 -1$), most of the time the scrambling pattern of those becomes similar to each other and hence reduce the level of security. Given below, in table 5, are the scrambled matrices of the above mentioned matrix A (equation (15)) for different $GFT_i$ $F(11)LT_i$ and $F(31)LT_i$.

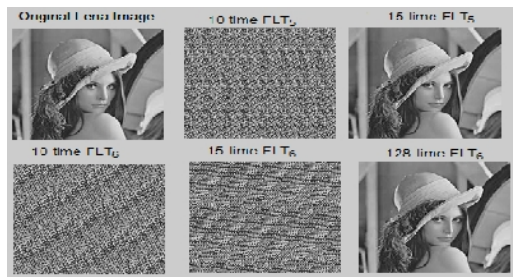

Figure 5. Lena image scrambled through $F(11)LT_5$ and $F(11)LT_6$

Table 5. Matrix A scrambled by different $FLT_i$ and $GFT_i$

| i | 1 | 2 | 3 | 4 | 5 | 6 | 7 | 8 |
|---|---|---|---|---|---|---|---|---|
| **Period of $F(11)LT_i$** | 8 | 6 | 6 | 6 | 8 | 3 | 2 | 3 |
| Scrambled A using $F(11)LT_i$ | 1 6 8 / 9 2 4 / 5 7 3 | 1 5 9 / 3 4 8 / 2 6 7 | 2 3 1 / 8 9 7 / 5 6 4 | 1 4 7 / 6 9 3 / 8 2 5 | 1 8 6 / 5 3 7 / 9 4 2 | 1 9 5 / 2 7 6 / 3 8 4 | 1 3 2 / 4 6 5 / 7 9 8 | 1 7 4 / 8 5 2 / 6 3 9 |
| | 1 4 7 / 3 6 9 / 2 5 8 | 1 4 7 / 9 3 6 / 5 8 2 | 3 1 2 / 6 4 5 / 9 7 8 | 1 6 8 / 3 5 7 / 2 4 9 | 1 4 7 / 3 6 9 / 2 5 8 | 1 4 7 / 9 3 6 / 5 8 2 | | 1 6 8 / 3 5 7 / 2 4 9 |
| | 1 9 5 / 8 4 3 / 6 2 7 | 1 3 2 / 7 9 8 / 4 6 5 | 1 2 3 / 7 8 9 / 4 5 6 | 1 3 2 / 7 9 8 / 4 6 5 | 1 5 9 / 6 7 2 / 8 3 4 | | | |
| | 1 3 2 / 7 9 8 / 4 6 5 | 1 9 5 / 2 7 6 / 3 8 4 | 2 3 1 / 5 6 4 / 8 9 7 | 1 7 4 / 8 5 2 / 6 3 9 | 1 3 2 / 7 9 8 / 4 6 5 | | | |
| | 1 8 6 / 5 3 7 / 9 4 2 | 1 7 4 / 5 2 8 / 9 6 3 | 3 1 2 / 9 7 8 / 6 4 5 | 1 8 6 / 2 9 4 / 3 7 5 | 1 6 8 / 9 2 4 / 5 7 3 | | | |
| | 1 7 4 / 2 8 5 / 3 9 6 | | | | 1 7 4 / 2 8 5 / 3 9 6 | | | |
| | 1 5 9 / 6 7 2 / 8 3 4 | | | | 1 9 5 / 8 4 3 / 6 2 7 | | | |
| **Period of $GFT_i$** | 8 | 6 | 2 | 6 | 8 | 3 | 2 | 3 |
| Scrambled A using $GFT_i$ | 1 4 7 / 5 8 2 / 9 3 6 | 1 8 6 / 2 9 4 / 3 7 5 | 1 9 5 / 4 3 8 / 7 6 2 | 1 3 2 / 8 7 9 / 6 5 4 | 1 7 4 / 9 6 3 / 5 2 8 | 1 6 8 / 3 5 7 / 2 4 9 | 1 5 9 / 7 2 6 / 4 8 3 | 1 2 3 / 6 4 5 / 8 9 7 |
| | 1 5 9 / 8 3 4 / 6 7 2 | 1 7 4 / 8 5 2 / 6 3 9 | | 1 2 3 / 5 6 4 / 9 7 8 | 1 5 9 / 8 3 4 / 6 7 2 | 1 7 4 / 8 5 2 / 6 3 9 | | 1 2 3 / 5 6 4 / 9 7 8 |
| | 1 8 6 / 3 7 5 / 2 9 4 | 1 3 2 / 7 9 8 / 4 6 5 | | 1 6 8 / 7 9 8 / 4 6 5 | 1 3 2 / 2 4 9 / 3 5 7 | | | |
| | 1 3 2 / 7 9 8 / 4 6 5 | 1 6 8 / 3 5 7 / 2 4 9 | | 1 2 3 / 6 4 5 / 8 9 7 | 1 3 2 / 7 9 8 / 4 6 5 | | | |
| | 1 7 4 / 9 6 3 / 5 2 8 | 1 4 7 / 6 9 3 / 8 2 5 | | 1 3 2 / 9 8 7 / 5 4 6 | 1 4 7 / 5 8 2 / 9 3 6 | | | |
| | 1 9 5 / 6 2 7 / 8 4 3 | | | | 1 9 5 / 6 2 7 / 8 4 3 | | | |
| | 1 6 8 / 2 4 9 / 3 5 7 | | | | 1 8 6 / 3 7 5 / 2 9 4 | | | |
| **Period of $F(32)LT_i$** | 8 | 3 | 2 | 3 | 8 | 6 | 2 | 6 |
| Scrambled A using $F(32)LT_i$ | 1 7 4 / 9 6 3 / 5 2 8 | 1 6 8 / 3 5 7 / 2 4 9 | 1 5 9 / 7 2 6 / 4 8 3 | 1 2 3 / 6 4 5 / 8 9 7 | 1 5 9 / 8 3 4 / 6 7 2 | 1 8 6 / 2 9 4 / 3 7 5 | 1 9 5 / 4 3 8 / 7 6 2 | 1 3 2 / 8 7 9 / 6 5 4 |
| | 1 5 9 / 8 3 4 / 6 7 2 | 1 7 4 / 8 5 2 / 6 3 9 | | 1 2 3 / 5 6 4 / 9 7 8 | 1 8 6 / 3 7 5 / 2 9 4 | 1 7 4 / 8 5 2 / 6 3 9 | | 1 2 3 / 5 6 4 / 9 7 8 |
| | 1 6 8 / 2 4 9 / 3 5 7 | | | | 1 3 2 / 7 9 8 / 4 6 5 | 1 3 2 / 7 9 8 / 4 6 5 | | 1 3 2 / 7 9 8 / 4 6 5 |
| | 1 3 2 / 7 9 8 / 4 6 5 | | | | 1 7 4 / 9 6 3 / 5 2 8 | 1 6 8 / 3 5 7 / 2 4 9 | | 1 2 3 / 6 4 5 / 8 9 7 |
| | 1 4 7 / 5 8 2 / 9 3 6 | | | | 1 9 5 / 6 2 7 / 8 4 3 | 1 4 7 / 6 9 3 / 8 2 5 | | 1 3 2 / 9 8 7 / 5 4 6 |
| | 1 9 5 / 6 2 7 / 8 4 3 | | | | 1 4 7 / 5 8 2 / 9 3 6 | | | |
| | 1 8 6 / 3 7 5 / 2 9 4 | | | | 1 6 8 / 2 4 9 / 3 5 7 | | | |

* Matrices of the last iterations are excluded from the list as these are same as the original matrix.

From the above table, it is clear that the maps with periods near the value $N^2$-1 (8 in this case, as N=3 for a 3x3 matrix) produces similar types of scrambling patterns, may be, with a different order. So, it is advisable not to pick those maps as the scramblers in order to improve the level of security of the encrypted message else a message scrambled by one map can be decrypted by another even if, the accurate map is not known. For example, both $GFT_1$ and $GFT_5$ in the above table 5 have the same maximum period (p= 8 =$N^2$ -1) and it can be noticed that both of them produce the same scrambling patterns with a different ordering. Therefore, a message scrambled through $GFT_1$ for a certain number of times, t< p, can safely be retrieved by $GFT_5$ by iterating it for some number of times between 1 and 8. But, maps with lower periods i.e., with period around N (2 and 3 in this case) have complete different patterns and can be considered for the purpose of higher security image scrambling. In case of images, since the value of N i.e., the size of the image generally is very high, a map with a period of N will be sufficient enough to be considered as a scrambler. Figure 5 shows the scrambling and retrieval of Lena image by two different FLTs.

Observation 2: The basic requirement for a map to be periodic is that the absolute value of the determinant of the matrix and the size of the image N should be co-primes. Hence, all unimodular matrices can be used as periodic scrambling maps (as ±1 and N (where NxN is the size of an image) are co-primes). Based on this principle, we can construct infinitely many 2x2 maps and if fact there exist as many as 24030 unimodular 2x2 maps between 0 and 99!  Though all of these maps may not provide unique scrabbling patterns and the scrambling patterns of many of those will be similar to each other still, any out of these 24030 unimodular maps can be used for our purpose.  The problem with this random selection of the map, of course, lies with the key transmission which requires the whole matrix to be transmitted along with the other keys whereas, in case of standard transforms, the requirement will be to transmit a single number indicating the starting term of the series to be transmitted.

Observation 3: When the transform period 'p' of a transform map TM is high and the number of iterations used during encryption 'q' is small, instead of iterating the encrypted message 'p-q' times during decryption, the image can be retrieved back much faster by simply iterating it for 'p' times through the inverse of TM as all the above mentioned maps are invertible maps. Figure 6 shows the encryption and decryption of the 128x128 gray scale Lena image using the map $F(11)LT_6 = \begin{pmatrix} 8 & 13 \\ 11 & 18 \end{pmatrix}$ as well as decryption through the Inverse$(F(11)LT_6) = \begin{pmatrix} 18 & -13 \\ -11 & 8 \end{pmatrix}$. The time elapsed is given in seconds (executed in MATLAB 7.x, 2.2 GHz processor) and it is clear that retrieval is much faster with inverse transformation.

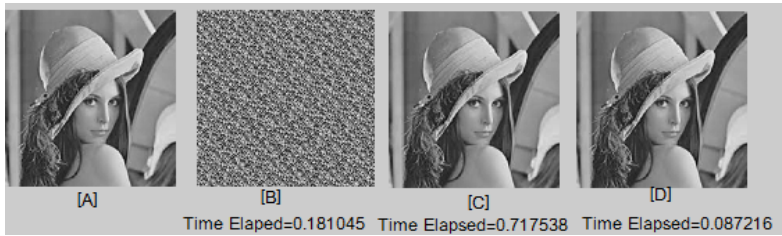

Figure 6.  [A]: Original Lena Image,
[B]: Lena image scrambled 20 times through $F(11)LT_6$
[C]: Applying 108 times $F(11)LT_6$ to B
[D]: Applying 20 times Inverse($F(11)LT_6$) to B

## 4. CONCLUSIONS

In this paper we have proposed a new spatial domain image scrambling map that can be used in various spatial domain image processing techniques of data hiding and secret communications such as Steganography and Watermarking and can increase the security of the hidden message.

We have experimentally proved that the transforms with higher periodicity may reduce the level of security as those can be decrypted by other maps even if the exact map is not known. Of course, a mathematical prove in this regard can help to decide which map and what iteration in that map will be more difficult to be decrypted without a valid key. With a valid key, the encrypted message can be decrypted by iterating it for p –p` times with the same map or with p` times by iterating it through the inverse map depending upon which value is smaller, whether p` or p-p` where, p is the period of the map and p` is the number of iterations performed in the encryption phase. It has also been proved through our experiments that the method is robust against statistical attacks such as cropping, format changes and introduction of different noises.

## REFERENCES

[1] http://en.wikipedia.org/wiki/Arnold%27s_cat_map
[2] V. I. Arnold; A. Avez (1968). Ergodic Problems in Classical Mechanics. New York: Benjamin.
[3] Ma, Z.G. and S.S. Qiu, 2003. "An image cryptosystem based on general cat map", J. China Inst. Commun., 24: 51-57.
[4] Kong, T. and Z. Dan, 2004. A new anti-Arnold transform algorithm. J. Software, 15: 1558-1564.
[5] Hong, C.Y. and W.G. Zou, 2005. "Digital image scrambling technology based on three dimensions Arnold transform and its period", J. Nanchang Univ. Nat. Sci., 29: 619-621.Wang,
[6] Z.H., 2006. "On the period of 2D "Random matrix scrambling transform and its application in image hiding", Chinese J. Comput., 29: 2218-2225.
[7] Yang, D.L., N. Cai and G.Q. Ni, 2006. "Digital image scrambling technology based on the symmetry of arnold transform", J. Beijing Inst. Technol., 15: 216-220.
[8] Minati Mishra, A.R. Routray, Sunit Kumar: "High Security Image Steganography with modified Arnold's cat map", IJCA, Vol.37, No.9:16-20, January 2012.
[9] Minati Mishra, Sunit Kumar and Subhadra Mishra: "Security Enhanced Digital Image Steganography Based on Successive Arnold Transformation", Advances in Intelligent and Soft Computing, 2012, Volume 167/2012, pp. 221-229, DOI: 10.1007/978-3-642-30111-7_21.
[10] Qi, D.X., J.C. Zou and X.Y. Han, 2000. "A new class of transform and its application in the image transform covering". Sci. China (Series E), 43: 304-312.
[11] Zou, J.C., R.K. Ward and D.X. Qi, 2004. "A new digital image scrambling method based on Fibonacci numbers". Proceedings of the International Symposium on Circuits and Systems, May 23-26, Vancouver, Canada, pp: 965-968.
[12] Zou, J.C., R.K. Ward and X.D. Qi, 2004. "The generalized fibonaci transformatios and application to image scrambling". Proceeding of the IEEE International Conference on Acoustic, Speech and Signal Processing, May 17-21, Canada, pp: 385-388.
[13] Li-Ping Shao, Zheng Qin, Hong-Jiang Gao and Xing-Chen Heng, 2008. "2D Triangular Mappings and Their Applications in Scrambling Rectangle Image", Information Technology Journal, 7: 40-47.